\documentclass[12pt]{article}
\usepackage[utf8]{inputenc}
\usepackage[a4paper]{geometry}
\usepackage{mathtools}
\usepackage{mathrsfs}
\usepackage{mathbbol}
\usepackage{amsthm}
\usepackage{amssymb}
\usepackage{tensor}
\usepackage{bm}
\usepackage{physics}

\usepackage{tikz}
\usetikzlibrary{decorations.markings}

\numberwithin{equation}{section}

\usepackage{setspace}
\usepackage{microtype} 
\usepackage{fancyhdr}
\usepackage[pdfusetitle]{hyperref}
\usepackage{xcolor}
\usepackage{lmodern}
\usepackage{enumitem}
\hypersetup{
	pdftitle={Localisation of Soft Charges, and Thermodynamics of Softly Hairy Black Holes},
	pdfauthor={Josh Kirklin, Malcolm Perry},
	colorlinks,
	linkcolor={red!50!black},
	citecolor={blue!50!black},
	urlcolor={blue!80!black},
	breaklinks=true
}
\usepackage[sorting=none,style=numeric-comp,maxbibnames=99,minbibnames=99]{biblatex}

\DeclareSymbolFontAlphabet{\amsmathbb}{AMSb}

\let\mathbb\amsmathbb

\newcommand{\scri}{\mathscr{I}}
\newcommand{\lie}{\mathcal{L}}
\newcommand{\hook}{\mathop\lrcorner}

\newenvironment{nalign}{ 
	\begin{equation}
	\begin{aligned}
}{
	\end{aligned}
	\end{equation}
	\ignorespacesafterend
}

\bibliography{refs}
\begin{document}
\def\UrlBreaks{\do\/\do-}

\begin{titlepage}
	\vspace*{1.35in}
	\begin{center}
		\parbox{\linewidth}{\centering \LARGE \bf Localisation of Soft Charges, and Thermodynamics of Softly Hairy Black Holes}
		
		\vspace*{2\baselineskip}
		
		{\large Josh Kirklin}
		
		\vspace*{1.5\baselineskip}
		
		{\em \large Department of Applied Mathematics and Theoretical Physics, Centre for Mathematical Sciences, University of Cambridge, Cambridge, UK}
		
		\let\thefootnote\relax\footnote{\url{jjvk2@cam.ac.uk}}
		
		\vspace*{2\baselineskip}
		
		{\large\bf Abstract}
	\end{center}
	\begin{quote}
		\small
		Large gauge transformations (LGT) in asymptotically flat space are generated by charges defined at asymptotic infinity. No method for unambiguously localising these charges into the interior of spacetime has previously been established. We determine what this method must be, and use it to find localised expressions for the LGT charges. Applying the same principle to the case of a charged black hole spacetime leads to angle-dependent generalisations of the Smarr formula and the first law of black hole mechanics, both of which have important thermodynamical implications. In particular, the presence of a heat current intrinsic to the event horizon is observed.
	\end{quote}
\end{titlepage}

\newpage

\tableofcontents

\newpage

\setcounter{footnote}{0}

\section{Introduction}

Consider a theory of fields in an asymptotically flat spacetime $\mathcal{M}$. In the covariant Hamiltonian approach to the analysis of such a theory, one must choose a $\partial\Sigma$ at which all Cauchy surfaces $\Sigma$ must have their boundary, and a set of boundary conditions at $\partial\Sigma$. If $\overline{\mathcal{M}}$ is the conformal compactification of $\mathcal{M}$ (and restricting to the case where $\partial\Sigma$ is connected), then the component of $\partial\Sigma$ at infinity can take one of three possible values:
\begin{itemize}
	\item Either $\partial\Sigma=i^0$, spacelike infinity, which is the singular point at infinite spacelike distance from all points in $\mathcal{M}$, or
	\item $\partial\Sigma=\scri^+_-$ or $\partial\Sigma=\scri^-_+$, the past/future endpoints of future/past null infinity respectively. Future/past null infinity are the unions of all points in $\partial\overline{\mathcal{M}}$ which are the future/past endpoints respectively of null curves originating in $\mathcal{M}$. 
\end{itemize}
Despite their proximity on a Penrose diagram, these three choices are not the same.

Historically, the most common choice has been $\partial\Sigma=i^0$. In some sense this is not particularly surprising; it is the most obvious immediate choice, especially in the absence of an understanding of the conformal structure of asymptotically flat space. However, from the point of view of the scattering problem, it is not the most helpful one. Spatial infinity is completely causally disconnected from the physical spacetime $\mathcal{M}$. In other words, an observer cannot exist at $i^0$. The only places early-time observations and late-time observations can be made are near $\scri^-\cup i^-$ and $\scri^+ \cup i^+$ respectively ($i^\pm$ are future/past timelike infinity, defined as the unions of all future/past endpoints of timelike curves from $\mathcal{M}$). Hence it makes the most sense in this context to pick $\partial\Sigma=\scri^+_-$ or $\partial\Sigma=\scri^-_+$. The classical scattering map can then be concretely realised as a bijection $S:Z^-\to Z^+$, where $Z^\pm$ are the phase spaces obtained by considering $\partial\Sigma=\scri^\pm_\mp$ respectively. 

It is not immediately clear how the system obtained by choosing $\partial\Sigma=\scri^\pm_\mp$ instead of $\partial\Sigma=i^0$ will differ. Certainly they will share most of their features. A recent fruitful line of research~\cites{Strominger:2013jfa,He:2014laa,Kapec:2014opa,He:2014cra,Strominger:2014pwa,He:2015zea,Kapec:2017tkm} has revealed that in gauge and gravity theories, there is at least one quality that the former have which is not shared by the latter. This is the existence of infinitely many more independent and physically significant degrees of freedom associated to gauge transformations. The charges generating these gauge transformations are known as soft charges, and have deep connections to classical memory effects and quantum mechanical soft theorems. Additionally, one particular scenario in which the quantum scattering problem has been difficult to tackle has been when $\mathcal{M}$ contains a black hole, and the newly observed existence of the soft charges has shed some light on this issue~\cite{Hawking:2016msc,Hawking:2016sgy}. For reviews on these topics, see~\cite{Strominger:2017zoo} or~\cite{Compere:2018aar}.

A key issue that has not yet been fully resolved is that of the localisation of these charges. One naturally initially derives expressions for the charges in terms of fields at infinity. One also interprets the charges as generating transformations of the fields at infinity. However, it is desirable to find expressions for the charges in terms of fields in the interior. For example if one is interested in the soft charges of a black hole, one would like to express these quantities in terms of fields on its event horizon. In~\cite{Hawking:2016sgy}, a specific gauge choice was made to extend the gauge transformation under consideration from infinity to the black hole horizon. Another approach has been to consider certain symmetries of the black hole horizon which are analogous to those at infinity, and to derive charges which generate these~\cite{Donnay:2015abr,Donnay:2016ejv}.

The first aim of this paper is to present a simple principle for the localisation of soft charges. Unlike the first approach above, the method is gauge invariant. It serves to provide a relation between the charges at infinity and those at the horizon obtained by the second approach above. It has a simple geometric interpretation.

The second aim of this paper is with regard to the thermodynamics of black holes. Consider a spacetime containing a black hole. The soft charges of this spacetime provide a notion of the energy, angular momentum, electric charge, etc.\ of the black hole at each angle on the celestial sphere, and it is natural to try to extend this so that one has a complete thermodynamical system at each angle. Using the localisation technique developed previously, we will generalise the laws of black hole mechanics so that they involve the soft charges. These generalised laws will consequently describe the thermodynamics at each angle, and lead to a natural definition of an angular entropy density. They will also reveal that the system at each angle is not closed, but is in thermal contact with the systems at neighbouring angles. We will find an expression for the resulting angular heat flux. This heat flux can also be viewed as existing on the horizon of the black hole.

The paper is laid out as follows. First, in Section \ref{Section: isolated}, we describe the Einstein-Maxwell description of isolated electromagnetic gravitational systems, and derive expressions for the soft charges of such systems. Section \ref{Section: localisation} then provides an explanation of our method for localising these expressions to the interior of the system. Next, in Section \ref{Section: thermodynamics}, we apply this technique to a stationary black hole. This allows us to obtain generalisations of Smarr's formula and the first law of black hole mechanics. Finally, we will close with some discussion on the results we have obtained, before suggesting future directions.

\section{Isolated electromagnetic gravitational systems}
\label{Section: isolated}
An electromagnetic gravitational system with metric $g_{ab}$ and electromagnetic potential $\mathcal{A}$ on a $4$-dimensional manifold $\mathcal{M}$ is described by the Einstein-Maxwell action $S=S_{\partial\mathcal{M}}+\int_{\mathcal{M}}L$. $S_{\partial\mathcal{M}}$ is a boundary term necessary to make the variational principle well defined, and we will not discuss it in detail here. $L$ is the Lagrangian density 4-form, and is given by\footnote{We ignore any matter contributions in this paper.}
\begin{equation}
	L = \frac1{16\pi G}\epsilon R + \frac1{2e^2}\mathcal{F}\wedge*\mathcal{F},
\end{equation}
where $R$ is the scalar curvature of $g_{ab}$, $\epsilon=\sqrt{- g}\dd[4]{x}$ is the volume form (with $g=\det g_{ab}$), $*$ is its associated Hodge star, and $\mathcal{F}=\dd{\mathcal{A}}$ is the electromagnetic field strength. $G$ and $e$ are coupling constants.

In this section we will analyse this system using the formalism described in~\cites{Wald:1993ki,Wald:1993nt,Iyer:1994ys,Wald:1999wa}, and consequently developed in many other papers.

The configuration space $C$ is the space of all possible smooth field configurations on $\mathcal{M}$. One can treat spaces of fields such as $C$ in a geometric way. The main step necessary to do this is to realise that a vector field on such a space corresponds to  a variation of the fields. A vector field specifies a different vector at each point, which in this context simply means that one can vary the fields in a field-dependent way. Equipped with this interpretation of the tangent bundle, one can construct and similarly interpret higher order tensors on $C$. For example, a field space 1-form is a linear functional of field variations.

Consider an infinitesimal field variation $\delta$, defined by $g_{ab}\to g_{ab}+\delta g_{ab}$, $\mathcal{A}\to \mathcal{A}+\delta\mathcal{A}$.\footnote{Notation: $\delta g_{ab} = \delta (g_{ab})$ is a tensor whose indices can be raised and lowered with the metric in the standard way. This means that (for example) $\delta g^{ab} = g^{ac}g^{bd}\delta g_{cd} = -\delta (g^{ab})$.} To linear order in $\delta g_{ab},\delta\mathcal{A}$, the corresponding variation of the Lagrangian density is given by $\delta L = \epsilon E^{ab}_{\text{Einstein}}\delta g_{ab} + E_{\text{Maxwell}}\wedge\delta\mathcal{A} + \dd{\theta}$, where
\begin{align}
	E^{ab}_{\text{Einstein}} &= -\frac1{16\pi G}\left(R^{ab}-\frac12 g^{ab} R\right) - \frac1{2e^2}\left(\mathcal{F}^{ca}\mathcal{F}\indices{_c^b} - \frac14 g^{ab}\mathcal{F}_{cd}\mathcal{F}^{cd}\right),\\
	E_{\text{Maxwell}} &= -\frac1{e^2}\dd{*\mathcal{F}}
\end{align}
are the left hand sides of the Einstein and Maxwell field equations $E^{ab}_{\text{Einstein}}=0, E_{\text{Maxwell}}=0$ respectively, and
\begin{equation}
	\theta(\delta) = \frac1{16\pi G}\epsilon_a(\nabla_b\delta g^{ab}-\nabla^a \delta g) + \frac1{e^2}\delta\mathcal{A}\wedge*\mathcal{F}
\end{equation}
is the Einstein-Maxwell presymplectic potential form. Here $\delta g=g^{ab}\delta g_{ab} = \delta (\ln g)$, and $\epsilon_a$ is defined such that $V^a\epsilon_a=V\hook\epsilon$. $\theta$ is a 3-form in spacetime, and at the same time a 1-form in field space, since it is a linear functional of the field variation $\delta$. 

Let $P$ be the level set in $C$ obeying the equations of motion. A field configuration $g_{ab},\mathcal{A}$ in $P$ is called on-shell. In a covariant Hamiltonian formalism, $P$ is used as the phase space of the theory (before gauge reduction). An on-shell field variation is one which takes on-shell configurations to on-shell configurations. In other words, as a vector field on $C$, it is tangent to $P$. The commutator of two on-shell field variations $\delta_1$ and $\delta_2$ leads to a third $\delta_{12}=[\delta_1,\delta_2]$, given by their field space Lie bracket.\footnote{The fact that $\delta_{12}$ is an on-shell field variation is a consequence of Frobenius' theorem.} 

We see that for any on-shell configuration, the variation of the Lagrangian density is an exact spacetime form $\delta L = \dd{\theta}$. Let $\delta_1,\delta_2$ be two on-shell field variations. By evaluating the trivial on-shell identity $0 = - \delta_1(\delta_2 L) + \delta_2(\delta_1 L) + \delta_{12}L$, one finds that $\dd{\omega}=0$, where 
\begin{align}
	\omega &= - \delta_1\big(\theta(\delta_2)\big) + \delta_2\big(\theta(\delta_1)\big) + \delta_{12} \theta \label{Equation: omega def}\\
	&= -\frac1{8\pi G} \Big[\delta_1(\epsilon_a g^{b[c})\delta_2\Gamma_{bc}^{a]} - \delta_2(\epsilon_a g^{b[c})\delta_1\Gamma_{bc}^{a]}\Big] + \frac1{e^2}\Big[\delta_1\mathcal{A}\wedge\delta_2(*\mathcal{F})-\delta_2\mathcal{A}\wedge\delta_1(*\mathcal{F})\Big]. \label{Equation: omega value}
\end{align}
We have used the identity $\nabla_b\delta g^{ab}-\nabla^a\delta g = 2g^{b[c}\delta \Gamma_{bc}^{a]}$ in order to write down the compact expression above.
$\omega$ is the Einstein-Maxwell presymplectic structure form. To obtain the Einstein-Maxwell presymplectic structure $\Omega$, one integrates $\omega$ over any Cauchy surface $\Sigma$. Since $\omega$ is closed as a spacetime form, it does not matter which Cauchy surface we pick, so long as $\partial\Sigma$ is fixed. At each point in the on-shell configuration space, $\Omega(\delta_1,\delta_2)$ is an antisymmetric bilinear functional of the two variations $\delta_1$ and $\delta_2$. One can thus view it as a 2-form on the on-shell configuration space $P$, and it can be verified that this 2-form is closed.

Note that $\Omega$ is only a \emph{pre}symplectic structure because it is not non-degenerate. Its degenerate directions correspond to redundancies in the description of the physical system provided by the metric $g_{ab}$ and gauge potential $\mathcal{A}$. One obtains the physical phase space by taking the quotient of $P$ by these degenerate directions.\footnote{Because $\Omega$ is closed, this is always possible, at least locally in field space.} $\Omega$ then naturally gives rise to a closed 2-form on the physical phase space, and this 2-form is the actual symplectic structure.

However, practically speaking, we do not usually need to actually carry out this reduction. For example, if we are interested in whether a certain variation $\hat\delta$ results in a physical change, it suffices to contract $\hat\delta$ into $\Omega$. If the resulting 1-form $\alpha(\delta) = \Omega(\hat\delta,\delta)$ vanishes, then $\hat\delta$ corresponds to a degenerate direction, and no physical change occurs. On the other hand, if $\alpha(\delta)$ is non-vanishing, $\hat\delta$ must have a component in a non-degenerate direction, and so a physical change is incurred.

Finally, note that if $\alpha$ is an exact form in field space, i.e.\ if we can write $\alpha(\delta) = \delta H$ for some function $H$, then we say that $\hat\delta$ is integrable, and that $H$ is the charge which generates $\hat\delta$. $H$ is invariant along the degenerate directions of the presymplectic structure, and so unambiguously gives rise to a function on the physical phase space.

\subsection{Large gauge transformations and soft charges}
A general gauge transformation in Einstein-Maxwell theory is the combination of a diffeomorphism, parametrised by a vector field $\chi$, and a Maxwell gauge transformation, parametrised by a function $\lambda$. Under this transformation, the metric and gauge potential infinitesimally transform as
\begin{align}
	g_{ab} &\to g_{ab} + \lie_\chi g_{ab} \nonumber\\
		&= g_{ab} + \nabla_a\chi_b + \nabla_b\chi_a,\\
	\mathcal{A} &\to \mathcal{A}+\lie_\chi\mathcal{A}+\dd{\lambda} \nonumber\\
		&= \mathcal{A}+\chi\hook\mathcal{F}+\dd{(\chi\hook\mathcal{A} + \lambda)}.
\end{align}
Denote this variation by $\delta_{\chi,\lambda}$. The objective of this section is to find which $\chi,\lambda$ give rise to physical $\delta_{\chi,\lambda}$. By `physical' we mean that the gauge transformation changes the physical state of the system. Their associated transformations are referred to as large gauge transformations (LGT). Non-physical gauge transformations are referred to as small. 

Naively one might expect that all gauge transformations must be small. However, this is not true when we are using a Cauchy surface with a non-empty boundary (or conformal boundary). The presence of such a boundary breaks gauge invariance.

To proceed, we will contract $\delta_{\chi,\lambda}$ into $\Omega$. We could use directly the expression for $\Omega$ given in \eqref{Equation: omega value}, but it turns out to be more convenient to return to the form of $\omega$ given by \eqref{Equation: omega def}. We shall set $\delta_1=\delta_{\chi,\lambda}, \delta_2=\delta$, and evaluate the result term by term.

The first term is $\delta_{\chi,\lambda}(\theta(\delta))$. The only part of $\theta(\delta)$ that transforms non-trivially under a Maxwell gauge transformation is $\delta\mathcal{A} \to \delta\mathcal{A}+\dd{\delta\lambda}$. Hence we have
\begin{equation}
	\delta_{\chi,\lambda}\big(\theta(\delta)\big) = \lie_\chi\theta(\delta) + \frac1{e^2}\dd{\delta\lambda}\wedge*\mathcal{F} = \dd{\left(\chi\hook\theta(\delta) +\frac1{e^2}\delta\lambda *\mathcal{F}\right)}+\chi\hook\dd{\big(\theta(\delta)\big)}
\end{equation}

The second term is $\delta(\theta(\delta_{\chi,\lambda}))$. First note that if we contract $\delta_{\chi,\lambda}$ into $\delta L = \dd{\big(\theta(\delta)\big)}$, we obtain\footnote{The Lagrangian density is Maxwell gauge invariant, so there is no contribution from $\lambda$ to the left hand side.}
\begin{equation}
	\dd{(\chi\hook L)} = \dd{\big(\theta(\delta_{\chi,\lambda})\big)},
\end{equation}
implying that $J_{\chi,\lambda}=\theta(\delta_{\chi,\lambda})-\chi\hook L$ is a closed 3-form. $J_{\chi,\lambda}$ is the (Hodge dual of) the Noether current associated to this gauge transformation. So long as there are no topological obstructions, we can therefore write $J_{\chi,\lambda} = \dd{Q_{\chi,\lambda}}$ for some 2-form $Q_{\chi,\lambda}$ -- the (Hodge dual of) the Noether charge for this gauge transformation. Indeed, we can take 
\begin{equation}
	Q_{\chi,\lambda} = \frac1{16\pi G}*\dd{\chi} + \frac1{e^2}(\chi\hook\mathcal{A}+\lambda)*\mathcal{F},
\end{equation}
where for notational simplicity we are using $\chi$ to mean both the 1-form $\chi^ag_{ab}\dd{x^b}$, and the vector $\chi^a\pdv{x^a}$. Hence we can write
\begin{equation}
	\delta\big(\theta(\delta_{\chi,\lambda})\big) = \dd{\delta (Q_{\chi,\lambda})}+ \delta(\chi\hook L).
\end{equation}
For the third term $\theta(\delta_{12})$, it helps to explicitly note what the action of $\delta_{12}=[\delta_{\chi,\lambda},\delta]$ is. We have
\begin{align}
	\delta_{\chi,\lambda}(\delta g_{ab}) - \delta(\delta_{\chi,\lambda}g_{ab}) &= \chi^c\partial_c\delta g_{ab}+\delta g_{ac}\partial_b\chi^c+\delta g_{bc}\partial_a\chi^c - \delta(\chi^c\partial_c g_{ab}+g_{ac}\partial_b\chi^c+g_{bc}\partial_a\chi^c) \nonumber\\
	&= -\delta\chi^c\partial_c g_{ab}-g_{ac}\partial_b\delta\chi^c-g_{bc}\partial_a\delta\chi^c = -\lie_{\delta\chi} g_{ab}\\
	\delta_{\chi,\lambda}(\delta \mathcal{A}) - \delta(\delta_{\chi,\lambda}\mathcal{A}) &= \chi\hook\delta\mathcal{F} + \dd{(\chi\hook\delta\mathcal{A}+\delta\lambda)} - \delta\big(\chi\hook\mathcal{F} + \dd{(\chi\hook\mathcal{A}+\lambda)}\big) \nonumber\\
	&=\delta\chi\hook\mathcal{F} + \dd{(\delta\chi\hook\mathcal{A})} = -\lie_{\delta\chi} \mathcal{A}.
\end{align}
Therefore, $\delta_{12}$ acts as an infinitesimal diffeomorphism along the vector field $-\delta\chi$. Hence we have
\begin{equation}
	\theta(\delta_{12}) = -\dd{Q_{\delta\chi,0}}-\delta\chi\hook L.
\end{equation} 

Putting the three terms together, we obtain
\begin{equation}
	\omega(\delta_{\chi,\lambda},\delta) = \dd{\left[-\chi\hook\theta(\delta) - \frac1{e^2}\delta\lambda*\mathcal{F}+\delta (Q_{\chi,\lambda}) - Q_{\delta\chi,0}\right]} - \chi\hook\dd{\big(\theta(\delta)\big)}+\delta(\chi\hook L) - \delta\chi\hook L.
\end{equation}
Using $\delta L = \dd{\theta}$, we see that the latter terms on the right hand side cancel, and we just get an exact form. To get $\Omega(\delta_{\chi,\lambda},\delta)$, we just need to integrate this form over the Cauchy surface $\Sigma$. The result is a boundary integral given by
\begin{equation}
	\Omega(\delta_{\chi,\lambda},\delta) = \int_{\partial\Sigma} \delta (Q_{\chi,\lambda}) - Q_{\delta\chi,\delta\lambda} - \chi\hook\theta(\delta).
	\label{Equation: gauge contracted Omega}
\end{equation}
Here we have combined $Q_{\delta\chi,0}+\frac1{e^2}\delta\lambda*\mathcal{F} = Q_{\delta\chi,\delta\lambda}$. 
Because $Q_{\chi,\lambda}$ is linear in $\chi$ and $\lambda$, $\Omega(\delta_{\chi,\lambda},\delta)$ must be independent of $\delta\chi$ and $\delta\lambda$. We can therefore choose the behaviour of $\delta\chi$ and $\delta\lambda$ at $\partial\Sigma$ in any way we like, and this will not reduce the set of independent physically significant transformations under consideration. It will however have an effect on whether or not these transformations are integrable.

We now have a condition for whether a gauge transformation is large or not. Namely, it is large if and only if \eqref{Equation: gauge contracted Omega} is non-vanishing. Additionally, if $\chi$ is tangent to $\partial\Sigma$, then we can set $\delta\chi=\delta\lambda=0$ and immediately obtain that $\int_{\partial\Sigma}Q_{\chi,\lambda}$ is the Hamiltonian charge generating the gauge transformation. The case where $\chi$ is not tangent to $\theta$ requires a slightly more detailed analysis, and it is usually only possible to make such transformations integrable by making use of supplementary boundary conditions.

\subsection{Charges of isolated systems}

An \emph{isolated} electromagnetic gravitational system is one for which $\mathcal{M},g_{ab}$ is asymptotically flat and the field strength $\mathcal{F}$ falls off at some physically sensible rate in the asymptotic region. In systems of this type it is possible to choose for the Cauchy surface to have its boundary at $\scri^+_-$ or $\scri^-_+$, the past/future endpoints of future/past null infinity respectively. We will focus on these systems and make this choice in what follows.

The requirement that the systems we are analysing be isolated / asymptotically flat means that we will need to impose some gauge-invariant boundary conditions on the metric and gauge field at infinity. These are necessary for the specification of the theory.

We will also make some gauge choices. A full analysis would require that these gauge choices could always be reached by doing a small gauge transformation. If this were not the case, then the gauge choice would put a restriction on the allowed physical states in which the system could be. It is not too hard to show that the Maxwell gauge we will take is non-restrictive, but it is less obvious that the same is true of the coordinates we will pick. In fact, there is evidence to the contrary, e.g.~\cite{Haco:2017ekf}. Nevertheless, this gauge choice is almost always made in similar analyses, and we will do the same, leaving the resolution of this important issue for later work.

We will focus on the case $\partial\Sigma=\scri^+_-$; the other choice $\partial\Sigma=\scri^-_+$ proceeds in a very similar manner. We pick retarded Bondi coordinates $(u,r,\Theta^A)$, in which constant $u$ surfaces are null, $g_{rA}=g_{rr}=0$ and $\det(g_{AB}/r^2)$ is a function of $\Theta^A$ alone. We can write the metric near future null infinity (which is reached by taking $r\to\infty$) as~\cites{Madler:2016xju,Hawking:2016sgy}
\begin{nalign}
	\dd{s}^2 = g_{ab}\dd{x^a}\dd{x^b}=& -\dd{u}^2 - 2\dd{u}\dd{r} + r^2\gamma_{AB}\dd{\Theta^A}\dd{\Theta^B} \\
	& +\frac{2m_b}{r}\dd{u}^2 + rC_{AB}\dd{\Theta^A}\dd{\Theta^B} + D^BC_{AB}\dd{u}\dd{\Theta^A} \\
	& +\frac1{16r^2}C_{AB}C^{AB}\dd{u}\dd{r}\\
	& +\frac{1}{r}\left(\frac43 N_A+\frac43u\partial_Am_b + \frac13C_{AB}D_CC^{BC}\right)\dd{u}\dd{\Theta^A} \\
	& +\frac14\gamma_{AB}C_{CD}C^{CD}\dd{\Theta^A}\dd{\Theta^B} \\
	& + \dots
	\label{Equation: retarded Bondi gauge}
\end{nalign}
The first line is the Minkowski metric. Later terms represent corrections to flat space. Constant $u,r$ surfaces have spherical topology. $C_{AB},N_A,m_b$ all depend on $u,\Theta^A$ only, and capital Latin letters are lowered and raised with the unit round metric on the sphere $\gamma_{AB}$ and its inverse $\gamma^{AB}$; its associated covariant derivative is $D_A$. $C_{AB}$ is traceless with respect to $\gamma_{AB}$. The fields $C_{AB},N_A,m_b$ are related to each other, and to the Maxwell field, by the Einstein equations.

$\scri^+_-$ is reached in these coordinates by considering a constant $u,r$ surface, taking $r\to\infty$, and then taking $u\to-\infty$.

For the Maxwell field we choose retarded radial gauge $\mathcal{A}_r=0$, $\mathcal{A}_u|_{\scri^+} = 0$, and boundary conditions such that we can write near future null infinity
\begin{equation}
	\mathcal{A} = \left(\frac1r E+O(r^{-2})\right)\dd{u} + (A_A+O(r^{-1}))\dd{\Theta^A},
	\label{Equation: retarded radial gauge}
\end{equation}
where $E,A_A$ are functions of $u,\Theta^A$ only.

We are assuming that all physical states can be put into the forms above. As a consequence we need now only consider those gauge transformations which preserve them. The diffeomorphisms which preserve \eqref{Equation: retarded Bondi gauge} are the Bondi-Metzner-Sachs (BMS) transformations~\cites{Bondi:1962px,Sachs:1962wk}. The components of a vector field $\zeta$ which generates a BMS transformation must take the following form at large $r$:
\begin{nalign}
	\zeta^u &= Z \equiv f+\frac12uD_A Y^A,\\
	\zeta^r &= -\frac12 r D_A Y^A + \frac12 D^2 Z  - \frac1{4r}(C^{AB}D_AD_BZ + 2D_AC^{AB}D_BZ) + O(r^{-2}),\\
	\zeta^A &= Y^A - \frac1rD^A Z + O(r^{-2}),
	\label{Equation: BMS generator}
\end{nalign}
where $f,Y^A$ depend only on $\Theta^A$, and $Y^A$ obeys the conformal Killing equation with respect to $\gamma_{AB}$, i.e.\ $D_AY_B+D_BY_A - \frac12 \gamma_{AB}D_CY^C=0$. The function $f$ is said to parametrise the `supertranslation' part of $\zeta$, and the vector $Y^A$ the `superrotation' part. A pure supertranslation is one with $Y^A=0$, and a pure superrotation is one with $f=0$. Note that $\zeta$ can only be exponentiated to a finite, non-singular diffeomorphism if $Y$ is a global conformal Killing vector on the 2-sphere. Nevertheless, when considering infinitesimal transformations, it is valid to allow $Y$ to take any value in the much larger space of general conformal Killing vectors.

The action of this BMS transformation on $\mathcal{A}$ is given by
\begin{equation}
	\lie_\zeta\mathcal{A} = O(r^{-1})\dd{u}+O(r^{-2})\dd{r}+\dd{\Theta^A}\big(Z\partial_u A_A + \partial_A(Y^BA_B) + O(r^{-1})\big).
\end{equation}
The conditions from \eqref{Equation: retarded radial gauge} on the $u,\Theta^A$ components of the gauge field are preserved by this transformation. However, the condition that $\mathcal{A}_r=0$ is not. We will need to combine the BMS transformation with an appropriate Maxwell gauge transformation to preserve this condition. The allowed gauge transformations are given by $\mathcal{A}\to\mathcal{A}+\dd{\tau}$, where
\begin{equation}
	\tau = \varepsilon +  \int\dd{r}\big(\zeta^a\mathcal{F}_{ar}+\partial_r(\zeta^a\mathcal{A}_a)\big) = \varepsilon + O(r^{-1}).
\end{equation}
$\varepsilon$ is any function that depends only on $\Theta^A$, and parametrises the Maxwell LGT.

In summary the remaining infinitesimal gauge transformations must have parameters of the above forms $\chi=\zeta$ and $\lambda=\tau$. We are now in a position to substitute our boundary conditions, gauge choices, and allowed residual gauge transformations into \eqref{Equation: gauge contracted Omega}. A fair amount of algebra later, one obtains
\begin{equation}
	\Omega(\delta_{\zeta,\tau},\delta) = \delta (H[f,Y,\varepsilon]) - H[\delta f, \delta Y, \delta\varepsilon] - T[f,Y],
\end{equation}
where
\begin{align}
	H[f,Y,\varepsilon] &= \int_{\scri^+_-}\dd[2]{\Theta}\sqrt{\gamma}\left[\frac{m_b}{4\pi G}f + \left(\frac{N_A}{8\pi G}+\frac{EA_A}{e^2}\right)Y^A + \frac{E}{e^2}\varepsilon\right],\\
	T[f,Y] &=\int_{\scri^+_-}\dd[2]{\Theta}\sqrt{\gamma} \left(f+\frac12uD_AY^A\right)\left(\frac{N_{BC}\delta C^{BC}}{16\pi G}+\frac{\partial_u A_A\delta A^A}{e^2}\right).
\end{align}
$N_{AB}=\partial_uC_{AB}$ is the Bondi news.

The usual step now is to assume that we have boundary conditions such that $T[f,Y]$ vanishes; the standard choice is that the Bondi news $N_{AB}$ and tangential components of the electric field $\partial_u A_A$ decay more quickly than $1/u$ as we approach $\scri^+_-$. We can then set $\delta f = \delta Y^A = \delta\varepsilon = 0$, and find that supertranslations, superrotations and Maxwell large gauge transformations are all integrable, and are generated by $H[f,Y,\varepsilon]$.

The problem with this is that the boundary conditions at $\scri^+_-$ are not preserved by all of these large gauge transformations. For example, one can calculate that a superrotation acts on the news as
\begin{equation}
	\delta N_{AB} = \lie_YN_{AB}-D_AD_BD_CY^C+\frac12\gamma_{AB}D^2D_CY^C.
\end{equation}
This only preserves the condition given above on the news if $Y^A$ is a global conformal KVF on the round sphere, but we want to be able to include \emph{all} superrotations, including those that are not global. Thus these charges and their resulting algebra will not be able to be exponentiated in a well-defined way.\footnote{The fact that non-global superrotations will be singular at certain points on the sphere is a separate, and much less serious, obstruction to exponentiation of the algebra.}

We will ignore this issue in this paper, and therefore will assume that the large gauge transformation charge with supertranslation parameter $f$, superrotation parameter $Y^A$, and Maxwell LGT parameter $\varepsilon$ of an asymptotically flat spacetime is given by $H[f,Y,\varepsilon]$.

There are a few special cases of which we should make note. We define
\begin{align}
	M[f] &= \int_{\scri^+_-}\dd[2]{\Theta}\sqrt\gamma\frac{m_b}{4\pi G}f, \\
	J[Y] &= \int_{\scri^+_-}\dd[2]{\Theta}\sqrt\gamma\left(\frac{N_A}{8\pi G} +\frac{EA_A}{e^2}\right)Y^A, \\
	Q[\varepsilon] &= \int_{\scri^+_-}\dd[2]{\Theta}\sqrt\gamma\frac{E}{e^2},
\end{align}
so that $H[f,Y,\varepsilon] = M[f]+J[Y]+Q[\varepsilon]$. $M=M[1]$ and $Q=Q[1]$ are the total mass and electric charge of the spacetime respectively. When $f$ is an $l=1$ harmonic, $M[f]$ gives some component of the total momentum of the spacetime. When $Y^A$ is a global conformal KVF on the round sphere (these form a six-dimensional space), $J[Y]$ gives some component of the total angular momentum and boost charge of the spacetime. We will call $M[f]$ the mass weighted by $f$, $J[Y]$ the angular momentum weighted by $Y^A$, and $Q[\varepsilon]$ the electric charge weighted by $\varepsilon$. By substituting delta functions into the arguments of these three functions, we get 
\begin{align}
	m(\Theta) &= {}^2\epsilon\frac{m_b}{4\pi G},\\
	j_A(\Theta) &= {}^2\epsilon\left(\frac{N_A}{8\pi G} + \frac{EA_A}{e^2}\right),\\
	q(\Theta) &= {}^2\epsilon\frac{E}{e^2},
\end{align}
where ${}^2\epsilon$ is the pullback of $\sqrt\gamma\dd[2]{\Theta}$ to $\scri^+_-$, and the right hand sides of these equations are each evaluated at the angle $\Theta$ on $\scri^+_-$.

One interpretation of the above results is that there are independent physical gauge degrees of freedom associated to each null generator of $\scri^+$. $m(\Theta),j_A(\Theta),q(\Theta)$ generate time translations, Lorentz transformations, and Maxwell gauge transformations on the null generator labelled by the angle $\Theta$. It is for this reason that we will call $m,j_A,q$ the angular densities of mass, A.M., and electric charge respectively. 

\section{Localisation of soft charges}
\label{Section: localisation}
The expressions found in Section \ref{Section: isolated} are all in terms of fields at infinity. However, if we want to discuss the soft charges of objects in the interior, we really want to be able to write down expressions in terms of fields near those objects. The objective of this section is to carry out this procedure of localisation. In particular, let $\tilde\Sigma$ be a surface such that $\scri^+_-$ is only one component of $\partial\tilde\Sigma$, and define $S = \partial\tilde\Sigma \setminus \scri^+_-$. We will write down the soft charges in terms of integrals over $S$.

\subsection{Maxwell LGT charge}
We initially focus on the soft electric charge, which can be written as
\begin{equation}
	Q[\varepsilon] = \frac1{e^2}\int_{\scri^+_-}\varepsilon*\mathcal{F}.
\end{equation}
By the Maxwell equations of motion $\dd{*\mathcal{F}}=0$, we have
\begin{equation}
	Q^{\tilde\Sigma}[\varepsilon]=\frac1{e^2}\int_{\tilde\Sigma} \dd{\varepsilon}\wedge*\mathcal{F} = Q[\varepsilon] - \frac1{e^2}\int_S \varepsilon*\mathcal{F}.
\end{equation}
Suppose that $Q^{\tilde\Sigma}[\varepsilon]$ vanishes. Then clearly $Q[\varepsilon]=\frac1{e^2}\int_S\varepsilon*\mathcal{F}$ is an expression for the soft charge associated to $\varepsilon$. This equality holds for all solutions of the Maxwell equations of motion, but more importantly the expressions for $Q[\varepsilon]$ as defined at $\scri^+_-$ and as defined at $S$ are completely physically equivalent in a Hamiltonian sense, in that they generate the same flow on phase space. In this way we have localised the charge $Q[\varepsilon]$ to $S$.

The simplest example is obtained by setting $\varepsilon=1$. $Q=Q[\varepsilon] = \frac1{e^2}\int_{\scri^+_-}*\mathcal{F}$ is then just the total electric charge of the spacetime. $\dd{\varepsilon}$ vanishes, so $Q^{\tilde\Sigma}[\varepsilon]=0$, and we find an equally valid expression for the total electric charge, $Q=Q^S=\frac1{e^2}\int_S*\mathcal{F}$.

We want to repeat this exercise for a more general choice of $\varepsilon$. In fact a similar kind of scenario will continue to arise during this paper. We will now lay out some machinery for application to the general case, before specialising to the electromagnetic LGT charges, and then other examples in later sections. 

Suppose we have an integral of the form $I[f]=\int_{\scri^+_-} f \beta$, where $f$ is a weight function on $\scri^+_-$, and $\beta$ is a closed $(n-2)$-form. Let $I^{\tilde\Sigma}[F]=\int_{\partial\tilde\Sigma} F \beta =\int_{\tilde\Sigma} \dd{F}\wedge\beta$, where $F=f$ on $\scri^+_-$. A sufficient condition for $I^{\tilde\Sigma}[F]=0$ is the vanishing of the pullback of $\dd{F}\wedge\beta$ to $\tilde\Sigma$. This can be written as 
\begin{equation}
	n_a(*\beta)^{ab}\partial_bF=0,
\end{equation}
where $n$ is a non-vanishing normal to $\tilde\Sigma$. In other words, $F$ need only be constant along integral curves of $n_a(*\beta)^{ab}$. If we choose $n=\dd{t}$ where $t$ is a level-surface function specifying $\tilde\Sigma$, this vector field is divergence-free. Hence its integral curves can only end at $\partial\tilde\Sigma$. Therefore the map $U_\beta$ which takes each point in $\partial\tilde\Sigma$ to the other end of the integral curve through that point is a well-defined involution of $\partial\tilde\Sigma$.\footnote{If the vector field vanishes at a point in $\partial\tilde\Sigma$, then there \emph{is} no integral curve through that point. We will ignore this issue. It is not really a problem, because at the points where the vector field vanishes there will be no contribution to $I[f]$ anyway.} We will make the assumption that $U_\beta(\scri^+_-)\cap\scri^+_-$ is empty, i.e.\ that each integral curve intersects $\scri^+_-$ no more than once. We then have $S_\beta \equiv U_\beta(\scri^+_-)\subset S$. Picking $F$ to be constant along integral curves, we can write
\begin{equation}
	I[f] = \int_{\scri^+_-}f\beta = \int_{S_\beta} (f\circ U_\beta)\beta = I^S[f].
\end{equation}
We will call the right hand side of the above equation the `localised' form of $I[f]$. Let $\beta|_{\scri^+_-}, \beta|_S$ be the pullbacks of $\beta$ to $\scri^+_-$ and $S$ respectively. Since $f$ is arbitrary in the above equation, this is really a relation between these two forms:
\begin{equation}
	\beta|_{\scri^+_-}=U_\beta^*\beta|_S.
\end{equation}
Note that the well-defined-ness of the right hand side is contingent on the smoothness of $U_\beta$, and this property is not guaranteed. We will ignore this issue. The right hand side is certainly well-defined where $U_\beta$ is smooth; we will treat it as a formal expression wherever this does not hold.

So consider now the soft electric charges. In the absence of matter $*\mathcal{F}$ is closed by the Maxwell equations. In this case the vector field along which the weight function should be constant is just the electric field $E^a=n_b\mathcal{F}^{ab}$. Assuming that $U_{*\mathcal{F}}(\scri^+_-)\cap\scri^+_-=\emptyset$, we can thus write down a localised form of the LGT charge
\begin{equation}
	Q^S[\varepsilon] = \frac1{e^2}\int_{S_{*\mathcal{F}}}(\varepsilon\circ U_{*\mathcal{F}})*\mathcal{F},
\end{equation}
or, in terms of angular charge densities $q=\frac1{e^2}*\mathcal{F}|_{\scri^+_-}$ and $q_S=\frac1{e^2}*\mathcal{F}|_S$,
\begin{equation}
	q=U_{*\mathcal{F}}^*q_S.
\end{equation}
This expression successfully localises an arbitrary soft electric charge to a finite subregion of spacetime. This localisation has a simple geometric interpretation -- the Maxwell gauge transformation parameter must be constant along electric field lines. This serves to demonstrate that electric field lines have an important role to play in the story of soft electric charges.

\begin{figure}
	\centering
	\begin{tikzpicture}[scale=0.7]
	\begin{scope}[rotate=20]
	\draw[very thick,fill=black!7] (0,0) circle (6);
	\begin{scope}[blue,dashed]
	\draw (0,2) .. controls (4.5,0.3) and (4.5,-0.3) .. (0,-2);
	\draw (0,2) .. controls (2.4,0.5) and (2.4,-0.5) .. (0,-2);
	\draw (0,2) .. controls (0.7,1) and (0.7,-1) .. (0,-2);
	\draw (0,2) .. controls (-4.5,0.3) and (-4.5,-0.3) .. (0,-2);
	\draw (0,2) .. controls (-2.4,0.5) and (-2.4,-0.5) .. (0,-2);
	\draw (0,2) .. controls (-0.7,1) and (-0.7,-1) .. (0,-2);
	
	\draw (0,2) .. controls (3.5,1.4) and (4.2,0.5) .. ({6*cos(10)},{6*sin(10)});
	\draw (0,2) .. controls (4,2) and (4.5,2.3) .. ({6*cos(30)},{6*sin(30)});
	\draw (0,2) .. controls (3.5,3) and (3.5,4.2) .. ({6*cos(50)},{6*sin(50)});
	\draw (0,2) .. controls (1,2.5) and (1.3,4) .. ({6*cos(70)},{6*sin(70)});
	\draw (0,2) -- (0,6);
	\draw (0,2) .. controls (-3.5,1.4) and (-4.2,0.5) .. ({-6*cos(10)},{6*sin(10)});
	\draw (0,2) .. controls (-4,2) and (-4.5,2.3) .. ({-6*cos(30)},{6*sin(30)});
	\draw (0,2) .. controls (-3.5,3) and (-3.5,4.2) .. ({-6*cos(50)},{6*sin(50)});
	\draw (0,2) .. controls (-1,2.5) and (-1.3,4) .. ({-6*cos(70)},{6*sin(70)});
	
	\draw (0,-2) .. controls (3.5,-1.4) and (4.2,-0.5) .. ({6*cos(10)},{-6*sin(10)});
	\draw (0,-2) .. controls (4,-2) and (4.5,-2.3) .. ({6*cos(30)},{-6*sin(30)});
	\draw (0,-2) .. controls (3.5,-3) and (3.5,-4.2) .. ({6*cos(50)},{-6*sin(50)});
	\draw (0,-2) .. controls (1,-2.5) and (1.3,-4) .. ({6*cos(70)},{-6*sin(70)});
	\draw (0,-2) -- (0,-6);
	\draw (0,-2) .. controls (-3.5,-1.4) and (-4.2,-0.5) .. ({-6*cos(10)},{-6*sin(10)});
	\draw (0,-2) .. controls (-4,-2) and (-4.5,-2.3) .. ({-6*cos(30)},{-6*sin(30)});
	\draw (0,-2) .. controls (-3.5,-3) and (-3.5,-4.2) .. ({-6*cos(50)},{-6*sin(50)});
	\draw (0,-2) .. controls (-1,2.-5) and (-1.3,-4) .. ({-6*cos(70)},{-6*sin(70)});
	
	\draw[thick,->,solid] (-3.5,1) -- (-5.7,0.74) node[midway,below\underline{}] {$E$};
	\end{scope}
	\begin{scope}[red,very thick,decoration={
		markings,
		mark=at position 0.5 with {\arrow{<}}}
	]
	\draw[postaction={decorate}] (0,2) .. controls (3.5,3) and (3.5,4.2) .. ({6*cos(50)},{6*sin(50)});
	\draw[postaction={decorate}] (0,2) .. controls (4,2) and (4.5,2.3) .. ({6*cos(30)},{6*sin(30)});
	\draw[postaction={decorate}] (0,2) .. controls (3.5,1.4) and (4.2,0.5) .. ({6*cos(10)},{6*sin(10)});
	\draw[postaction={decorate}] (0,-2) .. controls (3.5,-1.4) and (4.2,-0.5) .. ({6*cos(10)},{-6*sin(10)});
	\draw[postaction={decorate}] (0,-2) .. controls (4,-2) and (4.5,-2.3) .. ({6*cos(30)},{-6*sin(30)});
	\draw[postaction={decorate}] (0,-2) .. controls (3.5,-3) and (3.5,-4.2) .. ({6*cos(50)},{-6*sin(50)});
	\draw[postaction={decorate}] (0,-2) .. controls (1,-2.5) and (1.3,-4) .. ({6*cos(70)},{-6*sin(70)});
	\draw[postaction={decorate}] (0,-2) -- (0,-6);
	\draw[postaction={decorate}] (0,-2) .. controls (-1,2.-5) and (-1.3,-4) .. ({-6*cos(70)},{-6*sin(70)});
	\end{scope}
	
	\draw[very thick,fill=white] (0,2) circle (1);
	\draw[very thick,fill=white] (0,-2) circle (1);
	
	\begin{scope}[line width = 4pt, red]
	\draw ({6*cos(55)},{6*sin(55)}) arc (55:-115:6);
	\draw ({cos(25)},{sin(25)+2}) arc (25:-15:1);
	\draw ({cos(15)},{sin(15)-2}) arc (15:-140:1);
	\end{scope}
	\end{scope}
	\node at (-6,3) {$\scri^+_-$};
	\node at (-1,2) {$S^+$};
	\node at (0.4,-1.7) {$S^-$};
	\end{tikzpicture}
	\caption{Example of localisation of soft electric charge by propagation along electric field lines. This is a top-down view of the surface $\tilde\Sigma$. The outer boundary denotes $\scri^+_-$, and the inner boundaries $S^+$ and $S^-$ comprise $S=\partial\tilde\Sigma\setminus\scri^+_-$. The lines in the interior are the integral curves of the electric field $E$. Consider a soft charge with asymptotic support in the thick red region of $\scri^+_-$. Once localised, this will take the form of an integral with support in the thick red regions of $S^+$ and $S^-$. }
	\label{Figure: soft electric charge localisation}
\end{figure}
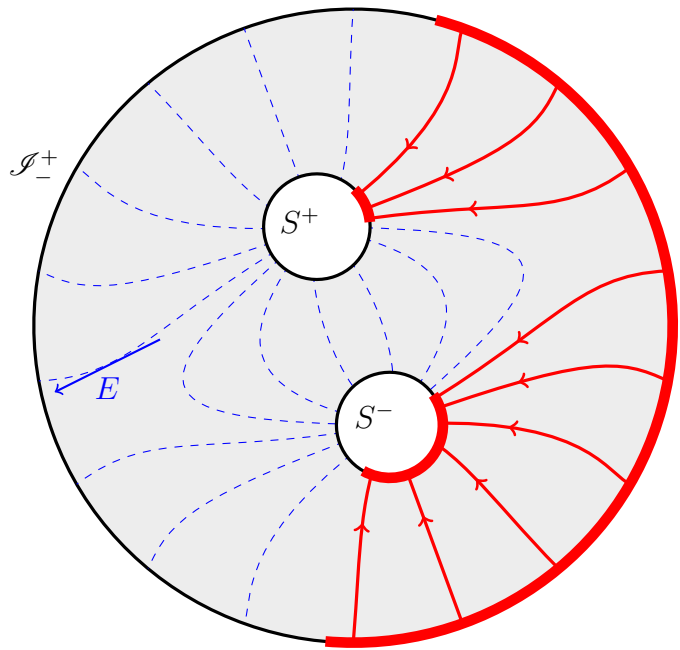

An example is provided in Figure \ref{Figure: soft electric charge localisation}.
There are a few noteworthy features of this example. For example, there are regions of $S$ that are completely inaccessible to the localisation procedure. These are those for which the appropriate electric field line both starts and ends on $S$. Also, the localisation map $U_{*\mathcal{F}}$ can be seen to be non-smooth at certain points, where its image jumps between $S^+$ and $S^-$. However, note that  at these points the electric field changes direction, and the charge density vanishes. Hence we do not need a localisation at these points anyway. It seems likely that a similar kind of thing happens in the generic case.

\subsection{BMS charge}
We now wish to carry out the same procedure for the BMS charges $M[f]$ and $J[Y^A]$. In order to do so we will make some quite restrictive assumptions about the spacetime we are dealing with. It seems likely that a construction can be found that does not make these assumptions. However, the assumptions hold in the main topic of interest (the study of stationary black holes), so we will use them in what follows.

First, suppose $\chi$ is a Killing vector of $g_{ab}$, and $\lambda$ is such that $\dd{\lambda}=-\lie_\chi\mathcal{A}$. We refer to such a $\chi,\lambda$ as a Killing pair. Then we have $\theta(\delta_{\chi,\lambda}) = 0$, since $\theta$ is linear in the field variations, and these just vanish. Using the formula $\theta(\delta_{\chi,\lambda}) = \dd{Q_{\chi,\lambda}} + \chi\hook L$, and the fact that the Ricci scalar $R$ vanishes whenever the equations of motion hold, we therefore have
\begin{align}
	0 &= \dd{Q_{\chi,\lambda}} + \frac1{2e^2}\chi\hook(\mathcal{F}\wedge*\mathcal{F}) \nonumber\\
	&= \dd{\left[Q_{\chi,\lambda} - \frac1{2e^2}\big((\chi\hook\mathcal{A}+\lambda)*\mathcal{F}+\mathcal{A}\wedge(\chi\hook*\mathcal{F})\big)\right]},
\end{align}
where the second line is most easily reached by repeated application of the magic formula $\lie_\chi\alpha=\chi\hook\dd{\alpha}+\dd{(\chi\hook\alpha)}$ and substitution of Maxwell's equation $\dd{*\mathcal{F}}=0$. Hence we discover that for each Killing pair $\chi,\lambda$ we have an associated exact 2-form given by the contents of the square brackets, explicitly $\dd{N[\chi,\lambda]}=0$ where
\begin{equation}
	N[\chi,\lambda] = \frac1{16\pi G}*\dd{\chi} + \frac1{2e^2}\big(\chi\hook(\mathcal{A}\wedge*\mathcal{F}) + \lambda*\mathcal{F}\big).
\end{equation}
In the case $\chi=0,\lambda=\text{constant}$, this reduces to Maxwell's equation. Note that this result is a special case of the generalised Noether theorem obtained in~\cite{Barnich:2000zw}.

The assumption we make to localise the supertranslation charge $M[f]$ is that spacetime is stationary with timelike KVF $k = \partial_u$, with both the metric and Maxwell field invariant under the action of $\lie_t$. It can then be shown that
\begin{equation}
	N[k,0]|_{\scri^+_-} = \epsilon\frac{m_b}{8\pi G}.
\end{equation}
Therefore, the supertranslation charge in the stationary case can be written as
\begin{equation}
	M[f] = 2\int_{\scri^+_-} f N[k,0].
\end{equation}
Similarly, to localise the superrotation charge $J[Y]$, we assume that spacetime is axially symmetric with rotational KVF $\psi$, and that we can write $\psi$ as a BMS transformation generating vector field of the form \eqref{Equation: BMS generator}, with $f=0$ and $Y^A = \psi^A$ a global conformal KVF on the sphere. We then find that
\begin{equation}
	N[\psi,0]|_{\scri^+_-} = \epsilon \left(\frac{N_A}{8\pi G}+\frac{EA_A}{e^2}\right)\psi^A.
\end{equation}
So, in the case that we can write $Y^A=h\psi^A$ for some function $h$, we can write the superrotation charge in the axially symmetric case as
\begin{equation}
	J[h\psi] = \int_{\scri^+_-} h N[\psi,0].
\end{equation}

To localise these charges, we can now just follow the same procedure as for the electric charge. We find
\begin{equation}
	M[f] = M^S[f] = 2\int_{S_{N[t,0]}}(f\circ U_{N[t,0]})N[t,0]
\end{equation}
and
\begin{equation}
	J[h\psi] = J^S[h\psi] = \int_{S_{N[\psi,0]}}(h\circ U_{N[\psi,0]})N[\psi,0].
\end{equation}
In terms of angular densities, we have
\begin{align}
	m &= U^*_{N[t,0]}m_S,\\
	j^A\psi_A &= U^*_{N[\psi,0]}j_S,
\end{align}
where $m_S, j_S$ are defined as the pullbacks of $2N[t,0], N[\psi,0]$ to $S$ respectively.

\section{Thermodynamics of soft hair}
\label{Section: thermodynamics}

The study of black hole thermodynamics was one of the original approaches to understanding semiclassical gravity. In the 1970s, a large amount of progress was made in this area. It was recognised that a set of four observations about black hole spacetimes could be viewed as being in close analogy with the four laws of thermodynamics~\cite{Bardeen:1973gs}. In particular, the entropy and temperature were in analogy to
\begin{equation}
	S \sim \frac{A}{\lambda},\qquad T \sim \lambda\frac{\kappa}{8\pi G},
\end{equation}
where $A$ is the surface area of the event horizon of the black hole, $\kappa$ is its surface gravity, and $\lambda$ is some constant. For the analogy to be exact, the black hole would have to radiate energy like any other hot body, and in the classical theory this cannot be true, since, by definition, no causal curve can be traced from the black hole region to future null infinity. However, taking quantum effects into account, Hawking famously showed that black holes do in fact radiate, and moreover that they radiate like a black-body at a temperature $T=\kappa/2\pi$~\cite{Hawking:1974rv}. This equation implies that $\lambda=4G$, and so the black hole's entropy should be given by $S=A/4G$.

A problem with this picture was almost immediately observed~\cite{Hawking:1976ra}. One considers a situation in which we have some cloud of matter in a pure quantum state, with many independent quantum numbers. One allows the matter to collapse to a black hole, which at late times one should expect to be approximately stationary. The black hole uniqueness and no-hair theorems~\cite{Mazur:2000pn} tell us that that this black hole is classified by only very few numbers: its mass, angular momentum, and charge. The black hole Hawking radiates quickly enough that it must eventually lose all energy and evaporate in finite time according to an observer. The spectrum of radiation is a probabilistic one parametrised only by the quantum numbers of the black hole. It would seem that the degrees of freedom that entered the black hole must be present in the radiation, but the limited parametrisation of the Hawking radiation cannot contain all of these degrees of freedom. The conclusion one is led to draw is that some degrees of freedom have completely disappeared! Information is lost; determinism is violated. The initially pure quantum state must evolve to a mixed one, and this is something that happens nowhere else in physics. This is one version of the black hole information paradox.

In~\cites{Hawking:2016msc,Hawking:2016sgy}, it was pointed out that in the context of the new soft charges, this argument may be flawed.
Any black hole spacetime may be mapped to a physically different black hole spacetime by the action of a large gauge transformation or large diffeomorphism. Furthermore, in a semiclassical approach, any classical stationary background spacetime may be used as a vacuum. Using a stationary black hole as the background and applying a spontaneous symmetry breaking argument, we observe that quantum black holes obtain a set of Goldstone modes. These are referred to as soft hair, and they invalidate the no-hair theorem in the quantum context. The authors of~\cite{Hawking:2016msc} conjecture that the soft hair will be sufficient to restore the information that is seemingly lost in black hole evaporation. Whether this is true is still a matter of debate, and we will not attempt to settle it here. For recent viewpoints, see~\cite{Mirbabayi:2016axw,Bousso:2017dny,Strominger:2017aeh}.

Nevertheless, there are still problems one can hope to solve in this context without running into too much controversy. A natural question to ask is whether one can obtain versions of the laws of black hole mechanics which respect the soft charges, and whether one can give these a thermodynamical interpretation. We will refer to these as the laws of black hole mechanics at every angle, and their derivation and exposition is the objective of this section. 

It is worth noting that an appropriate zeroth law and second law have already been shown to hold at every angle. The conventional zeroth law is the statement that the surface gravity of a stationary black hole is constant over the horizon. This trivially implies that the surface gravity is pointwise constant, which is the zeroth law at every angle. The conventional second law is the statement that the area of the event horizon of the black hole cannot decrease. A second law at every angle would then have to be that the expansion of each null generator of the horizon is non-negative. But showing that such a statement holds is a step in most proofs of the traditional second law. See for example~\cite[Lemma 9.2.2]{Hawking:1973uf}.

The third law is much less concrete than the other three. One way of stating it is: it is impossible for the surface gravity of an initially non-extremal (i.e.\ non-vanishing surface gravity) black hole to be reduced to zero everywhere on the horizon in a finite number of steps. It seems natural to guess that the generalisation to every angle should be one of two possibilities: it is impossible for the surface gravity of a black hole to be reduced from a non-zero value to zero at either a single point on the horizon, or in the neighbourhood of any point on the horizon, in a finite number of steps. Since the traditional third law has not been rigorously proven, we will not attempt to carry out a proof of a third law at every angle at this stage. We will only comment that if one can be shown to be true, then it seems likely that the other can too.

It remains to generalise the first law of black hole mechanics to one concerning charges at every angle. The traditional first law describes a relation that must hold if we perturb a black hole by a small amount. In Einstein-Maxwell theory, this is
\begin{equation}
	\delta M = \frac{\kappa}{8\pi G}\delta A - \Omega\delta J - \Phi\delta Q,
\end{equation}
where $M$ is the mass of the black hole, $\Omega$ its angular velocity, $J$ its angular momentum, $\Phi$ its electric potential, $Q$ its electric charge, $\kappa$ its surface gravity, and $A$ its area. It is reasonable to expect that one can find a similar identity that relates instead these quantities at every angle. In a sense, the above law is an integral one: it relates quantities that are obtained by integrating over a time slice of the event horizon. The first law at every angle that we obtain is in this sense a differential one, relating quantities that are defined pointwise on the event horizon. Schematically, it takes the form
\begin{equation}
	\delta m = \frac{\kappa}{8\pi G}\delta a + \nabla \cdot l - \Omega \delta j - \Phi \delta q,
\end{equation}
where $m,a,j,q$ are densities that integrate to their capitalised counterparts, and $\nabla\cdot l$ is the divergence of a vector field $l$ tangential to the horizon that depends linearly on the field perturbations. As we will discuss, $l$ has a natural thermodynamical interpretation as a heat flow tangential to the horizon. Note that upon integration over the horizon this divegence term disappears, and we obtain again the conventional first law.

In this section we will focus on asymptotically flat stationary spacetimes containing a single non-extremal electrically charged black hole. In these spacetimes we have access to a stationary KVF $k$ and a rotational KVF $\psi$. We will normalise these such that $k$ has unit norm and the orbits of $\psi$ have period $2\pi$ at infinity. As above, we will write the electrostatic potential of the black hole relative to infinity as $\Phi$, and its angular velocity as $\Omega$. The vector field $\xi=k+\Omega\psi$ is the KVF that generates the event horizon.

\subsection{Smarr's formula}
\begin{figure}
	\centering
	\begin{tikzpicture}[scale=0.6,thick]
	\draw (0,0) -- (6,6) -- (12,0) -- (6,-6) -- cycle;
	\draw[blue, dashed] (0,0) .. controls (3,0) and (9,3) .. (12,0) node[midway, above] {$\Sigma$};
	\node[blue, left] at (0,0) {$\mathcal{S}$};
	\node[blue, above right] at (12,0) {$\scri^+_-$};
	\node[above left] at (3,3) {$\mathcal{H}^+$};
	\node[below left] at (3,-3) {$\mathcal{H}^-$};
	\node[above right] at (9,3) {$\scri^+$};
	\node[below right] at (9,-3) {$\scri^-$};
	\end{tikzpicture}
	\caption{The domain of dependence of the surface $\tilde\Sigma$ in a stationary black hole spacetime chosen such that $\partial\tilde\Sigma = \scri^+_- \cup \mathcal{S}$.}
	\label{Figure: charged black hole}
\end{figure}
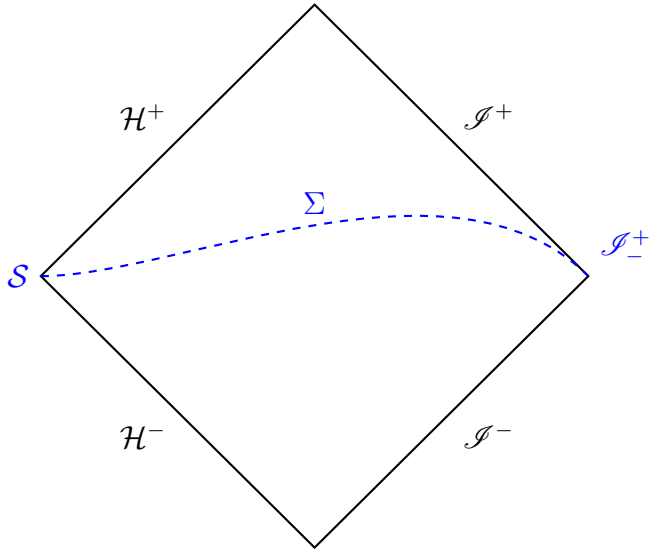

Before obtaining the first law, we will warm up with a generalised version of Smarr's formula. Consider the conserved 2-form $N \equiv N[\xi,\Phi]$, and let $\tilde\Sigma$ be a surface with boundary given by the disjoint union of $\scri^+_-$ and $\mathcal{S}$, the bifurcate 2-surface where the past event horizon $\mathcal{H}^-$ and future event horizon $\mathcal{H}^+$ meet. Figure \ref{Figure: charged black hole} depicts this scenario.

At $\scri^+_-$, $N$ pulls back to
\begin{equation}
	N|_{\scri^+_-} = \frac12m + \Omega \psi^Aj_A + \frac12\Phi q,
\end{equation}
and at $\mathcal{S}$ it pulls back to
\begin{equation}
	N|_{\mathcal{S}} = \frac{\kappa}{8\pi G} a,
	\label{Equation: Wald entropy density}
\end{equation}
where $\kappa$ is the surface gravity of the event horizon, and $a$ is the induced area element on the horizon. Hence, integrating $N$ over $\partial\tilde\Sigma$, one finds
\begin{equation}
	\frac12M+\Omega J+\frac12\Phi Q = \frac{\kappa}{8\pi G} A,
	\label{Equation: original Smarr}
\end{equation}
where $M$ is the mass of the black hole, $J$ is its angular momentum, $Q$ is its electric charge, and $A$ is its area. This is the Smarr formula.

\eqref{Equation: original Smarr} is an equation that only applies to the angular zero modes of the large gauge and large diffeomorphism charges. One can generalise it to one that has an angular dependence by using the localisation method established previously. Let $z$ be some function on the sphere. Then we have
\begin{equation}
	\frac12M[z]+\Omega J[z\psi]+ \frac12\Phi Q[z] = \int_{\scri^+_-}zN = \int_{\mathcal{S}_N}(z\circ U_N)N.
\end{equation}
Using \eqref{Equation: Wald entropy density}, and defining the weighted black hole area
\begin{equation}
	A[z]=\int_{\mathcal{S}_N}(z\circ U_N)a,
\end{equation}
one obtains
\begin{equation}
	\frac12 M[z] + \Omega J[z\psi] + \frac12\Phi Q[z] = \frac{\kappa}{8\pi G} A[z].
\end{equation}
This is an angular generalisation of the Smarr formula. One can also express it in terms of charge densities if we set $z$ to a delta function. We then obtain
\begin{equation}
	\frac12 m + \Omega \psi^Aj_A + \frac12\Phi q = \frac{\kappa}{8\pi G} U_N^*a.
\end{equation}

\subsection{The first law}
To obtain the first law, we will need a conserved 2-form that depends linearly on field variations. To find one, we consider now \eqref{Equation: gauge contracted Omega}, replicated below for convenience, when $\chi,\lambda$ is a Killing pair. 
\begin{equation*}
	\Omega(\delta_{\chi,\lambda},\delta) = \int_{\partial\Sigma} \delta (Q_{\chi,\lambda}) - Q_{\delta\chi,\delta\lambda} - \chi\hook\theta(\delta).
\end{equation*}
The left hand side clearly vanishes, since the presymplectic structure is linear in the variations. Thus in this case the integrand on the right hand side is a closed form. Explicitly, $\dd{T[\chi,\lambda]}=0$, where
\begin{equation}
	T[\chi,\lambda] = \delta(Q_{\chi,\lambda}) - Q_{\delta\chi,\delta\lambda} - \chi\hook\theta(\delta).
\end{equation}
So again consider a stationary charged black hole spacetime, and in particular let us choose the Killing pair $\chi,\lambda=\xi,\Phi$ as in the previous section. Let $\tilde\Sigma$ be as in the previous section, and define $T\equiv T_{\xi,\Phi}$. At $\scri^+_-$, $T$ pulls back to
\begin{equation}
	T|_{\scri^+_-} = \delta m + \Omega\psi^A\delta j_A + \Phi\delta q
\end{equation}
and at $\mathcal{S}$ it pulls back to
\begin{equation}
	T|_{\mathcal{S}} = \frac{\kappa}{8\pi G}\delta a
\end{equation}
Therefore if we integrate $\dd{T}$ over $\tilde\Sigma$, we obtain the celebrated first law of black hole mechanics in its standard form,
\begin{equation}
	\delta M + \Omega \delta J + \Phi \delta Q = \frac{\kappa}{8\pi G} \delta A.
\end{equation}

To find an angle-dependent first law, we can proceed in the usual way, but the vector field $n_a(*T)^{ab}$ is a little difficult to deal with. Our solution is to start by splitting up $T$.

Primes will denote varied fields:
\begin{equation}
	g'_{ab} = g_{ab}+\delta g_{ab}, \qquad \mathcal{A}' = \mathcal{A}+\delta \mathcal{A}.
\end{equation}
More generally primes will denote quantities derived using the primed fields. One can write $T = M-N$, where $N$ is defined as before, and
\begin{multline}
	M = N' + \frac1{2e^2}\Big(\chi\hook(\mathcal{A}'\wedge(*\mathcal{F})'-\mathcal{A}\wedge*\mathcal{F}) + \lambda\big((*\mathcal{F})'-*\mathcal{F}\big)\Big) \\
	- \chi\hook\big(\frac1{16\pi G}\epsilon_a\nabla_b(g'^{ab}-g^{ab}g_{cd}g'^{cd})+\frac1{e^2}(\mathcal{A}'-\mathcal{A})\wedge*\mathcal{F}\big)
\end{multline}
We have shown previously that $\dd{N}=0$. We also have that $\dd{M}=\dd{T}+\dd{N}=0$. Therefore, each of these forms is individually conserved. Because $T$ is linear in the field variations, $M$ and $N$ are infinitesimally close to each other. Furthermore $M$ and $N$ generically have non-zero parts that do not depend on the field variations. Therefore, $M$ and $N$ are much larger than their difference. This will become useful in what follows.

The left hand side of the generalised first law will take the form $\int_{\scri^+_-}zT$, where $z$ is some function. Define two functions $v$ and $w$ with the property that $v=w=z$ at $\scri^+_-$. Then we have
\begin{equation}
	\int_{\scri^+_-}zT = \int_{\scri^+_-}vM - \int_{\scri^+_-}wN.
\end{equation}
Now we will localise each integral on the right hand side individually. We get
\begin{equation}
	\int_{\scri^+_-}zT = \int_{\mathcal{S}_M} (v\circ U_M) M - \int_{\mathcal{S}_N}(w\circ U_N)N.
\end{equation}
Using a delta function for $z$ we get an expression in terms of densities
\begin{equation}
	T = U_M^*M- U_N^*N = U_N^* \big((U_M\circ U_N^{-1})^*M - N\big)
\end{equation}
Note that $U_M\circ U_N^{-1}$ is a diffeomorphism of $\mathcal{S}$. Since the difference between $N$ and $M$ is much smaller than either, it is safe to make the assumption that this diffeomorphism is infinitesimally close to the identity. Let it be characterised by a vector field $\hat{l}$ tangent to $\mathcal{S}$. $\hat{l}$ is linear in the field variations $\delta g_{ab},\delta \mathcal{A}$, but \emph{not} in a local fashion. We then have
\begin{equation}
	U_M^*M- U_N^*N = U_N^* \big(M + \lie_{\hat{l}}M - N\big) = U_N^*\big(T + \dd{(\hat{l}\hook N)}\big).
\end{equation}
But $N,T$ at $\mathcal{S}$ are just given by $\frac{\kappa}{8\pi G} a$ and $\frac{\kappa}{8\pi G}\delta a$ respectively. Therefore we obtain
\begin{equation}
	\delta m + \Omega \psi^A \delta j_A + \Phi \delta q = \frac\kappa{8\pi G} U_N^*\left[\delta a + \dd{(\hat{l}\hook a)}\right].
\end{equation}
One can expand the right hand side to find
\begin{equation}
	\delta m +\Omega \psi^A \delta j_A + \Phi\delta q = \frac\kappa{8\pi G}\delta(U_N^*a) + \frac\kappa{8\pi G} U_N^*\dd{(l\hook a)},
	\label{Equation: densitised first law}
\end{equation}
where $l$ is the vector field on the horizon generating the diffeomorphism $U_{N+P}U_N^{-1}$, and
\begin{equation}
	P = M - N' = \frac1{2e^2}\delta\big(\chi\hook(\mathcal{A}\wedge*\mathcal{F}) + \lambda*\mathcal{F}\big)
	- \chi\hook\left(\frac1{16\pi G}\epsilon_a(\nabla_b\delta g'^{ab}- \nabla^a\delta g)+\frac1{e^2}\delta\mathcal{A}\wedge*\mathcal{F}\right).
\end{equation}
Equation \eqref{Equation: densitised first law} is the first law of black hole mechanics at every angle. The right hand side contains a variation of the horizon area density, but also a term that appears to correspond to a horizon surface current $l$. This horizon surface current could equally well be interpreted as a surface current on the celestial sphere, by pulling it back through the map $U_N$.

Finally, note that if we integrate \eqref{Equation: densitised first law} over $\scri^+_-$ against a weight function $f$, we get this generalisation of the first law in integral form:
\begin{equation}
	\delta M[f] + \Omega \delta J[f\psi] + \Phi \delta Q[f] = \frac\kappa{8\pi G} \delta A[f] + \frac\kappa{8\pi G} \int_{\mathcal{S}_N} l(f\circ U_N) a.
\end{equation}
Note that if we set $f=1$, the rightmost term vanishes, and we just get back the first law in its usual form. The conventional first law is thus just one of the infinity of first laws provided by the above expression. 

\section{Discussion}
\label{Section: discussion}
In this paper we have proposed a simple method for the localisation of soft charges to the interior of a spacetime. We have also obtained a set of laws governing the soft charges of an asymptotically flat spacetime containing a black hole. The first three are:
\emph{\begin{enumerate}
	\setcounter{enumi}{-1}
	\item The surface gravity of a stationary black hole has vanishing gradient.
	\item A perturbation to a stationary black hole obeys \eqref{Equation: densitised first law}.
	\item The expansion along each null generator of the horizon is non-negative.
\end{enumerate}}
The third law we conjecture to have two possible forms:
\emph{\begin{enumerate}
	\setcounter{enumi}{2}
	\item It is impossible to reduce the surface gravity at any point (strong) / in the neighbourhood of any point (weak) on the horizon from a positive value to zero in a finite number of steps.
\end{enumerate}}

The original four laws of black hole mechanics are widely believed to arise from the thermodynamics of the microscopic physics of a near equilibrium black hole. The Bekenstein-Hawking entropy $S=A/4G$ strongly suggests that the microscopic states are in some way distributed over the black hole horizon. Therefore it seems reasonable to hope that this generalisation of the laws of black hole mechanics, which applies to each point on the horizon individually, has the potential to shed some new light on the microscopic degrees of freedom, which in this context are the soft hairs at each angle.

The above laws suggest a natural generalisation of black hole temperature and entropy, that should be expected to hold near equilibrium. Let $x$ be a point in the horizon. We propose that the entropy density $s(x)$ and temperature $t(x)$ of the black hole at $x$ should be given by
\begin{equation}
	s(x) = \frac{a(x)}{4G},\qquad t(x) = \frac{\kappa(x)}{2\pi}.
	\label{Equation: local entropy density and termperature}
\end{equation}
The non-negative expansion of the horizon implies that this definition of entropy density obeys the second law of thermodynamics. It would be of interest to compare the angular Hawking spectrum of a near equilibrium black hole temperature with the above value. 

Equations \eqref{Equation: local entropy density and termperature} could have been guessed without the above analysis, but the rightmost term in \eqref{Equation: densitised first law} suggests another, less obvious, part to this analogy. We propose that the 1-form 
\begin{equation}
	l(x)\hook \frac{a(x)}{4G}
\end{equation}
provides a natural candidate for the heat current of the horizon at the point $x$ for an approximately stationary black hole. This describes how energy is exchanged between the microscopic degrees of freedom of the black hole (i.e.\ the soft hair), and so it should hopefully provide some insight on how these are coupled together. The heat current is derived directly from the Wald-Noether charge density $N$ and the presymplectic potential form $\theta$. These are both intimately related to the information content of the spacetime, and this makes this definition particularly appealing. Of interest is the fact that the heat current appears to be constructed in a non-local manner from fields outside of the black hole. This follows from the appearance in its definition of maps that propagate along the integral curves of certain vector fields, and perhaps reflects the non-local behaviour that any quantum theory of gravity is believed to exhibit.

Besides the rather open-ended goal of exploring the consequences of the thermodynamical interpretation of the above laws, there are many directions in which this work could be taken in the future. We list a few below.

The maps that propagate along integral curves played a key role in the localisation of soft charges. This suggests that if one formulates theories with these charges in such a way that the integral curves are given an explicit role, then it may be possible to obtain some new insights. This would be interesting to explore.

It would be of use to understand the connection (if any) between the present work and the study of bit threads~\cite{Freedman:2016zud}. Both invoke ideas of divergence-free vector fields, and 1-to-1 maps between degrees of freedom and the flow lines of these vector fields.

The localisation of the gravitational soft charges described in this paper only works for some soft charges, and only in spacetimes which permit Killing fields. One should try to generalise the method so that Killing fields are not required.

The angular momentum term in the first law is $\Omega \psi^A\delta j_A$. It only has something to say about a certain component of the angular momentum density, namely the component in the direction of the rotational Killing field $\psi$. It might be worth exploring whether the other component of the angular momentum density has a role to play in black hole thermodynamics.

It would be useful to provide concrete examples of the applicability of the generalised first law. For example, it is known that throwing an asymmetric configuration of matter into a stationary black hole results in a change in its soft hairdo~\cite{Hawking:2016sgy}. The configuration of the spacetime after this process is a perturbation of the initial configuration, and therefore obeys the first law. This should be checked. In a similar vein, numerical simulations of perturbed black holes should obey the first law, and this might be worth testing.

Finally, one should evaluate the right hand side of the first law in the case of the Kerr-Newman black hole. The resulting explicit expression for $l$ may contain some interesting information.

\section*{Acknowledgements}
I would like to thank Malcolm Perry for many helpful discussions, David Skinner and Harvey Reall for some good advice, and Kelley Kirklin for some useful comments. I am also appreciative of the hospitality of the physics department at Harvard, where part of this work was carried out. This work was supported by a grant from STFC, and also grants from DAMTP and Clare College.

\raggedright
\printbibliography

\end{document}